\documentclass[runningheads]{llncs}

\usepackage{cite}
\usepackage{amsmath,amssymb,amsfonts}
\usepackage{mathdots}
\usepackage{algorithmicx}
\usepackage{graphicx}
\usepackage{caption}
\usepackage{subcaption}
\usepackage{textcomp}
\usepackage[dvipsnames]{xcolor}
\usepackage{orcidlink}
\usepackage{hyperref}
\usepackage{todonotes}
\usepackage{pgfplots}
\usepackage{pgf}
\usepackage{graphicx}
\usepackage{svg}
\usepackage{multirow}
\usepackage{bbm}
\usepackage{tikz}
\usetikzlibrary{quantikz2}

\usepackage[english]{babel}
\usepackage{algorithm2e}
\RestyleAlgo{ruled}
\usepackage[noend]{algpseudocode}
\usepackage{dsfont}
\usepackage{listings}

\usepackage[capitalise]{cleveref}

\newcommand{\probgate}[2]{\gate{#1}\gategroup[1,steps=1,style={rounded
         corners,fill=gray!20, inner
         xsep=2pt},background,label style={label
         position=below,anchor=north,yshift=-0.2cm}]{#2}}

\usepackage[T1]{fontenc}
\usepackage{graphicx}
\pgfplotsset{compat=1.18}
\begin{document}
\title{Optimization Framework for Reducing Mid-circuit Measurements and Resets}
%
%
\author{Yanbin Chen\orcidID{0000-0002-1123-1432} \and
Innocenzo Fulginiti\orcidID{0000-0001-8818-9626} \and
Christian B.~Mendl\orcidID{0000-0002-6386-0230}}
\authorrunning{Y. Chen, I. Fulginiti, et al.}
%
\institute{
School of CIT, Technical University of Munich, Garching 85748, Germany
\email{\{yanbin.chen, innocenzo.fulginiti, christian.mendl\}@tum.de}
}
\maketitle              
\begin{abstract}
The paper addresses the optimization of dynamic circuits in quantum computing, with a focus on reducing the cost of mid-circuit measurements and resets. 
We extend the probabilistic circuit model (PCM) and implement an optimization framework that targets both mid-circuit measurements and resets. To overcome the limitation of the prior PCM-based pass, where optimizations are only possible on pure single-qubit states, we incorporate circuit synthesis to enable optimizations on multi-qubit states. With a parameter $n_{pcm}$, our framework balances optimization level against resource usage.
We evaluate our framework using a large dataset of randomly generated dynamic circuits. Experimental results demonstrate that our method is highly effective in reducing mid-circuit measurements and resets. In our demonstrative example, when applying our optimization framework to the Bernstein-Vazirani algorithm after employing qubit reuse, we significantly reduce its runtime overhead by removing all of the resets.
\end{abstract}
\section{Introduction}\label{sec:intro}
Dynamic circuits, characterized by the inclusion of mid-circuit measurements and/or resets (operators that deterministically bring the qubit state to $\ket{0}$), are increasingly important in the design and implementation of quantum algorithms. Unlike static circuits, which rely solely on pre-defined quantum operations, dynamic circuits enable real-time feedback and conditional operations based on measurement outcomes \cite{ibm_mid_circ_meas_available_2021, nation_ibm_howtomidcircmeas_2021, ibm_dynamic_circuit}.
This adaptability greatly expands the potential of quantum algorithms, enabling dynamic approaches to tackle challenges in error mitigation, optimization problems, and intricate computational processes \cite{chiaverini2004realization, levine2009quantum, dong2022ground, kissinger2019universal, herman2023quantum}.
However, one of the challenging issues with dynamic circuits is the cost of performing mid-circuit measurements and resets \cite{PhysRevLett.127.100501, lubinski2022advancing, ella2023quantumclassical}. This difficulty arises due to several reasons.
First, mid-circuit measurements interrupt the coherent evolution of the quantum system, requiring precise synchronization between measurement operations and subsequent quantum gate execution \cite{Graham_2023, vazquez2024scalingquantumcomputingdynamic}.
Second, mid-circuit measurements involve interactions with classical systems to process outcomes in real-time, introducing latency that can degrade the overall performance of dynamic circuits \cite{lubinski2022advancinghybridquantumclassicalcomputation, C_rcoles_2021}. This latency becomes especially problematic in scenarios requiring fast feedback for classically controlled quantum operations.
Third, mid-circuit measurements are sensitive to noise, as the act of measuring qubits exposes the quantum system to external perturbation \cite{PhysRevApplied.10.034040, hashim2024quasiprobabilisticreadoutcorrectionmidcircuit, gaebler2021suppression}. Qubit resets, on the other hand, also pose technological challenges \cite{Zhou_2021, gehér2024resetresetquestion, Ding_2025}. Therefore, extra hardware support and error mitigation techniques are needed to counteract these effects. 

Given the challenges and costs associated with mid-circuit measurements and resets, optimizations that reduce mid-circuit measurements and resets for dynamic circuits are highly beneficial.
One such optimization pass introduces the concept of probabilistic gates, which,  by static analysis, captures the runtime behavior of mid-circuit measurements \cite{chen2024reducingmidcircuitmeasurementsprobabilistic}. This pass eventually replaces a subset of mid-circuit measurements in dynamic circuits with probabilistic sub-circuits. 
In this pass, however, the analysis may reveal possibilities of mid-circuit measurement reduction only in cases where measurement is performed on a pure state that is statically known. This significantly limits its ability to uncover optimization opportunities.
Besides, this pass does not reduce resets in the circuit. 
In addition, although the idea of this optimization pass is promising, it lacks an existing implementation and any evaluation demonstrating its effectiveness. 

We implement
the optimization pass for reducing mid-circuit measurements proposed in \cite{chen2024reducingmidcircuitmeasurementsprobabilistic}, and building on it, we develop an extended optimization framework, where we extend both the PCM and the quantum constant propagation (QCP), the static analysis technique used in \cite{chen2024reducingmidcircuitmeasurementsprobabilistic}. This framework not only broadens the scope of optimizations to include mid-circuit measurements performed on qubits entangled with the rest of the system, but also introduces the capability to reduce resets. Furthermore, the framework offers a parameter $n_{pcm}$ to trade-off between the level of optimization and the computational resources required, enabling adaptable and efficient optimization strategies. Then, we conduct a comprehensive evaluation of our framework using a large dataset of randomly generated dynamic circuits, demonstrating its effectiveness in reducing mid-circuit measurements and resets. Notably, when applied to circuits employing the technique of qubit reuse, our method is able to achieve substantial reductions in runtime overhead by eliminating the majority of, and in our demonstrative case, all of the resets introduced during applying the qubit reuse.
Our implementation is publicly available at \url{https://github.com/i2-tum/pcm-optimization-tool}.

\section{Preliminaries}\label{sec:prelim}
This manuscript presumes that readers have a foundational understanding of quantum computing. For a comprehensive introduction to the subject, we recommend consulting the following references: \cite{nielsen_QC_2012, rieffel2000introduction, kaye2006introduction}.  
In this section, we outline the notations that will be used throughout the paper.


\subsection{Static and dynamic circuits} \label{subsec:static-dynamic-circuit}
In the context of this manuscript, a \emph{static circuit} is a quantum circuit without any mid-circuit measurements and resets. On the contrary, a \emph{dynamic circuit} is a circuit with at least one mid-circuit measurement or reset. 
As mentioned in \cref{sec:intro}, in dynamic circuits, mid-circuit measurements and resets are major contributors to quantum run-time overhead. To distinguish the runtime overhead and static overhead, by \emph{static operation}, we refer to any quantum gate. On the contrary, a \emph{dynamic operation} refers to a mid-circuit measurement or reset. 


\subsection{Quantum constant propagation (QCP)} \label{subsec:QCP}
Quantum constant propagation is a symbolic execution technique designed to identify and eliminate superfluous controls and gates in circuits by taking advantage of static information propagated from their initial contexts \cite{chen_QCP_2023}. In this manuscript, it is assumed that the initial qubit states of circuits are all $\ket{0}$. 
QCP efficiently tracks quantum states by grouping entangled qubits, using a specially tailored union-table.
QCP maintains polynomial efficiency regardless of the number of qubits or circuit depth by introducing the parameter $n_{max}$, an upper limit on the size of entanglement groups it tries to track. When an entanglement group exceeds this limit, QCP stops tracking its details and assigns it the value $\top$, meaning no known static information.


\subsection{Probabilistic circuit model (PCM)} \label{subsec:PCM}
A \emph{probabilistic gate} $U(p)$, graphically denoted as follows, gives a stochastic scheme for compiling a gate:
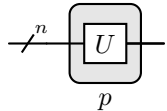
\begin{equation*}
    \begin{quantikz}[row sep=0.1cm]
        &\qwbundle{n}& \probgate{U}{$p$} & \qw
\end{quantikz}
\end{equation*}
Specifically, $U(p)$ is compiled to $U$ at probability $p$, and to the identity transformation $I$ at probability $1 - p$ \cite{10821051}.
A circuit that contains such probabilistic gates is a \emph{probabilistic circuit}. Each execution (or "shot") of a probabilistic circuit requires the compilation of probabilistic gates into specific non-probabilistic gate sequences. This process is guided by the probabilistic strategy, where the outcome of each probabilistic gate is determined at compile-time based on its given probability distribution. Consequently, the probabilistic circuit is transformed into a deterministic circuit tailored for that specific shot/execution.


\begin{example}
    When compiling and executing the probabilistic circuit \cref{fig:ex-pc-a}, one of the four possible circuits in \cref{fig:ex-pc-d} gets executed for each shot.

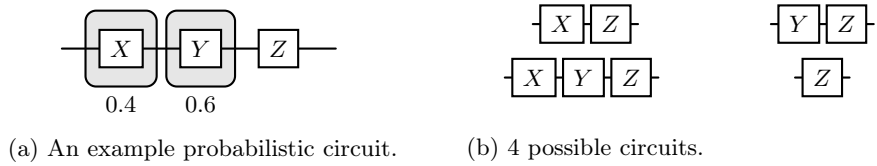
\begin{figure}[h]
  \centering
  \begin{tabular}[c]{ccc}
    \multirow{2}{*}[0pt]{
    \begin{subfigure}[t]{0.5\textwidth}
        \centering
        \begin{quantikz}
        &\probgate{X}{$0.4$}& \probgate{Y}{$0.6$}&\gate{Z}&
        \end{quantikz}
        \vspace{0pt}
        \caption{An example probabilistic circuit.}
        \label{fig:ex-pc-a}
    \end{subfigure}
}&
   \begin{subfigure}[t]{0.2\textwidth}
        \centering
        \begin{quantikz}[column sep=0.1cm]
        &\gate{X}&\gate{Z}&
        \end{quantikz}
    \end{subfigure}&
    \begin{subfigure}[t]{0.2\textwidth}
        \centering
        \begin{quantikz}[column sep=0.1cm]
        &\gate{Y}&\gate{Z}&
        \end{quantikz}
    \end{subfigure}\\
    &
    \begin{subfigure}[t]{0.3\textwidth}
        \centering
        \begin{quantikz}[column sep=0.1cm]
        &\gate{X}& \gate{Y}&\gate{Z}&
        \end{quantikz}
        \vspace{5.5pt}
        \caption{$4$ possible circuits.}
        \label{fig:ex-pc-d}
    \end{subfigure}&
     \begin{subfigure}[t]{0.2\textwidth}
        \centering
        \begin{quantikz}[column sep=0.1cm]
        &\gate{Z}&
        \end{quantikz}
    \end{subfigure}\\
  \end{tabular}    
  \caption{Compiling and executing an example probabilistic circuit. 
  }
  \label{fig:ex-pc}
\end{figure}

\end{example}

\subsection{PCM-based circuit optimization} \label{subsec:PCM-opt}
The PCM briefed in \cref{subsec:PCM} is introduced to construct an optimization pass that reduces the number of mid-circuit measurements for dynamic circuits \cite{chen2024reducingmidcircuitmeasurementsprobabilistic}. This pass uses QCP (see \cref{subsec:QCP}) to collect static information in the input dynamic circuit. Then, this information is used to replace mid-circuit measurements with probabilistic sub-circuits, thus decreasing quantum run-time overhead at a cost of a small amount of computation at compile-time. 

\begin{remark}
Importantly, when a mid-circuit measurement is replaced, its outcome is not lost; instead, it is precomputed at compile time according to the corresponding probability distribution and stored in the same classical wire as the original measurement outcome. This ensures that the program’s semantics remain unchanged while eliminating mid-circuit measurements.
\end{remark}

\begin{example}
When applying the PCM-based pass to \cref{ex:PCM-opt-a}, QCP is first performed to gather static information propagated from the initial context. As illustrated by \cref{ex:PCM-opt-b}, the particularly useful information is that the top qubit before the measurement is $\ket{+}$ and the bottom qubit before the reset is $\ket{+}$. With these static information, this pass replaces the mid-circuit measurement on the top qubit and the reset on the bottom qubit by probabilistic sub-circuits, as shown in \cref{ex:PCM-opt-c}. In this example, the pass successfully reduces the run-time overhead by removing all mid-circuit measurements and resets in the input circuit by introducing the compile-time cost to deal with the probabilistic gates.

\begin{figure}
    \centering
    \begin{subfigure}[t]{0.21\textwidth}
\begin{quantikz}[row sep=0.4cm, column sep=0.11cm]
    \lstick{\ket{0}}&\gate{H}&\meter{}\wire[d][1]{c}&\gate[3]{U}&  \\
    \lstick{\ket{0}}&        &\gate{Y}              &             &  \\
    \lstick{\ket{0}}&\gate{H}&\push{\ket{0}}        &             &
\end{quantikz}
    \vspace{3pt}
    \caption{Input circuit.}
    \label{ex:PCM-opt-a}
    \end{subfigure}
    ~
    \begin{subfigure}[t]{0.23\textwidth}
\begin{quantikz}[row sep=0.4cm, column sep=0.2cm]
    \lstick{}&\gate{H}\slice{\ket{+0+}}&\meter{}\wire[d][1]{c}&\gate[3]{U}&  \\
    \lstick{}&                 &\gate{Y}              &             &  \\
    \lstick{}&\gate{H}         &\push{\ket{0}}        &             &
\end{quantikz}
    \vspace{0pt}
    \caption{QCP analysis.}
    \label{ex:PCM-opt-b}
    \end{subfigure}
    ~
    \begin{subfigure}[t]{0.3\textwidth}
\begin{quantikz}[row sep=0.4cm, column sep=0.25cm]
    \lstick{}&\gate{H}&\gate{H}&\probgate{X}{$0.5$}&\ctrl{1}&\gate[3]{U}&  \\
    \lstick{}&& &             &\gate{Y} &            &  \\
    \lstick{}&\gate{H}&\gate{H}        &&&&
\end{quantikz}
    \vspace{3pt}
    \caption{The optimized circuit.}
    \label{ex:PCM-opt-c}
    \end{subfigure}
\caption{An example of applying the PCM-based pass on a dynamic circuit.}
\label{fig:PCM-ex}
\end{figure}
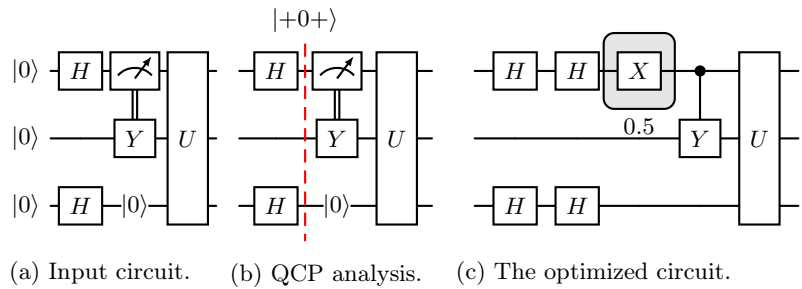
    \label{ex:PCM-opt}
\end{example}

\subsection{Quantum state preparation}
The goal of \emph{state preparation} is for an $n$-qubit state $\ket{\psi}$ to construct a circuit $C_{\ket{\psi}}$ such that $C_{\ket{\psi}}\ket{0}^{\otimes n} = \ket{\psi}$.
Many methods have been developed to achieve efficient and accurate state preparation \cite{zylberman2024efficient, zhang2022quantum, sun2023asymptoticallyoptimalcircuitdepth, rosenthal2023querydepthupperbounds}.
Procedures of state preparation have been integrated into modern quantum computing toolchains. For instance, Qiskit provides the $\texttt{StatePreparation}$ class. 
We only consider ancilla-free methods and assume we have access to a procedure that, given any state, generates a circuit that prepares the state, as detailed in \cref{def:state-prep}.
\begin{definition}[State preparation procedure]
    Given an $n$-qubit state $\ket{\psi}$, $SP_{\ket{\psi}}$ is a procedure generating a static circuit of which both the depth and gate count are bounded by $\mathcal{O}(2^n)$ \footnote{This asymptotic is a modest one compared to the latest results in state preparation, but it is already good enough for discussion in this manuscript.} that prepares the state $\ket{\psi}$, i.e., $SP_{\ket{\psi}}\ket{0}^{\otimes n} =  \ket{\psi}$.
\label{def:state-prep}
\end{definition}

\begin{definition}[Size of state]
    Given an $n$-qubit state $\ket{\psi}$, with its expansion in terms of the computational basis given by $\ket{\psi} = \sum_{i=0}^{2^n-1} \alpha_i \ket{i}$, the size of $\ket{\psi}$, denoted by $|\ket{\psi}|$, is given by $|\ket{\psi}| = \sum_{i=0}^{2^n-1}\mathbbm{1}(\alpha_i)$, i.e, the number of non-zero coefficients, where the indicator function $\mathbbm{1}(\cdot)$ is defined by:
    $
    \mathbbm{1}(x) :=
    \begin{dcases}
        0 & \text{if } x = 0 \\
        1 & \text{otherwise} \\
    \end{dcases}
    $
\label{def:size-state}
\end{definition}

\section{Method}\label{sec:methods}
In \cref{subsec:extended-PCM} and \cref{subsec:improved-PCM-pass}, we present the extensions to the PCM model and our optimization framework to address the challenges previously identified, namely, the inability to optimize mid-circuit measurements on entangled states and the need to optimize mid-circuit resets. \Cref{subsec:ext-QCP} details how we enhance QCP to support our extended framework. Since the optimization framework relies on the QCP to extract static information about quantum states, improving the QCP enables the detection of more static information, which, in turn, unlocks greater opportunities for circuit optimization through the pass.

\subsection{Extended PCM}\label{subsec:extended-PCM}

For conciseness of notation, we first extend the notation of probabilistic gates to \cref{def:ext-pg}. In \cref{ex:compile-ext-pg}, we provide an example of compiling and executing a probabilistic gate in the extended notation. In \cref{ex:relation-pg-to-ext-pg}, we see how the extended definition relates to its original definition.
\begin{definition}[Extended probabilistic gate]
    An extended probabilistic gate $\mathbf{G}_{\mathbb{P}}[(U_1, p_1), \dots, (U_k, p_k)]$, where for $\forall i \in \{1, \dots k\}$ $U_i$ is a quantum operator and $0\le p_i\le 1$, and $\sum^{k}_{1}p_i = 1$. The graphical notation is shown as follows:
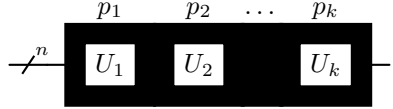
\begin{equation*}
    \begin{quantikz}
&\qwbundle{n}&\gate{U_1}\gategroup[1,steps=1,style={fill=black, inner
     sep=3.7pt}, background]{$p_1$}& \gate{U_2}\gategroup[1,steps=1,style={fill=black,inner
     sep=3.7pt}, background]{$p_2$} & \gategroup[1,steps=1,style={fill=black, inner
     sep=3.7pt}, background]{$\dots$}  &\gate{U_k}\gategroup[1,steps=1,style={fill=black,inner
     sep=3.7pt}, background]{$p_k$}&
\end{quantikz}
\end{equation*}
\label{def:ext-pg}
\end{definition}

\begin{example}
    The circuit \cref{fig:ex-ext-pc-a} contains a $S$ gate and a single extended probabilistic gate $\mathbf{G}_{\mathbb{P}}[(H \otimes I, 0.1), (I \otimes X, 0.2), (Z\otimes Z, 0.3), (X \otimes Y, 0.4)]$.
    When it is compiled, one of the $4$ possible static circuits is generated as shown in \cref{fig:ex-compile-ext-pg-circ}.
\begin{figure}[h]
  \centering
  \begin{tabular}[c]{ccc}
    \multirow{2}{*}[20pt]{
    \begin{subfigure}[t]{0.5\textwidth}
        \centering
        \begin{quantikz}[column sep=0.35cm]
        &\gate{S}&\gate{H}\gategroup[2,steps=1,style={fill=black, inner
         sep=3.7pt}, background]{$0.1$}& \gategroup[2,steps=1,style={fill=black,inner
         sep=3.7pt}, background]{$0.2$} &\gate{Z} \gategroup[2,steps=1,style={fill=black, inner
         sep=3.7pt}, background]{$0.3$}  &\gate{X}\gategroup[2,steps=1,style={fill=black,inner
         sep=3.7pt}, background]{$0.4$}& \\
         &&&\gate{X}&\gate{Z}&\gate{Y}&
        \end{quantikz}
        \vspace{0pt}
        \caption{An extended probabilistic circuit.}
        \label{fig:ex-ext-pc-a}
    \end{subfigure}
}&
   \begin{subfigure}{0.2\textwidth}
       \begin{quantikz}[row sep = 0.15cm, column sep=0.1cm]
            &\gate{S}& \gate{H}  & \\ 
            &&           &
        \end{quantikz}
        \caption{$C_1$ at probability $0.1$.}
        \label{fig:ex-ext-pc-b}
    \end{subfigure}&
    \begin{subfigure}{0.2\textwidth}
       \begin{quantikz}[row sep = 0.01cm, column sep=0.1cm]
            & \gate{S}& & \\ 
            & &\gate{X} &
        \end{quantikz}
        \caption{$C_2$ at probability $0.2$.}
        \label{fig:ex-ext-pc-c}
    \end{subfigure}\\
    &
    \begin{subfigure}{0.2\textwidth}
       \begin{quantikz}[row sep = 0.05cm, column sep=0.1cm]
            & \gate{S}&\gate{Z} &   \\ 
            & &\gate{Z}  &
        \end{quantikz}
        \caption{$C_3$ at probability $0.3$.}
        \label{fig:ex-ext-pc-d}
    \end{subfigure}&
     \begin{subfigure}{0.2\textwidth}
       \begin{quantikz}[row sep = 0.05cm, column sep=0.1cm]
            & \gate{S}&\gate{X}   & \\ 
            && \gate{Y} &
        \end{quantikz}
        \caption{$C_4$ at probability $0.4$.}
        \label{fig:ex-ext-pc-e}
    \end{subfigure}\\
  \end{tabular}    
  \caption{An example of compiling and executing a probabilistic circuit in the extended notation \cref{def:ext-pg}. When running the probabilistic circuit \cref{fig:ex-ext-pc-a}, one of the possible circuits, i.e., \cref{fig:ex-ext-pc-b}, \cref{fig:ex-ext-pc-c}, \cref{fig:ex-ext-pc-d}, \cref{fig:ex-ext-pc-e}, is executed conforming to the given probability distribution in \cref{fig:ex-ext-pc-a}. }
  \label{fig:ex-compile-ext-pg-circ}
\end{figure}
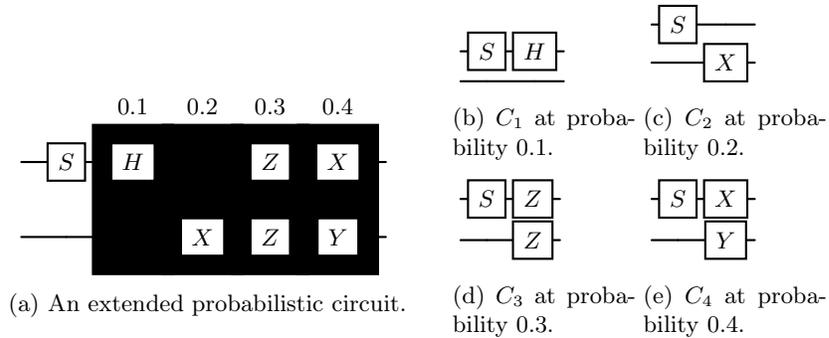
\label{ex:compile-ext-pg}
\end{example}

\begin{example}
    Given a probabilistic gate $U(p)$, since it compiles to $U$ at probability $p$ and to $I$ at probability $1 - p$, we can rewrite it in the extended notation as $\mathbf{G}_{\mathbb{P}}[(U, p), (I, 1 - p)]$.
\label{ex:relation-pg-to-ext-pg}
\end{example}

\subsection{Improved PCM-based optimization pass}\label{subsec:improved-PCM-pass}
Based on the extended PCM presented in \cref{subsec:extended-PCM}, we now present our method to extend the PCM-based pass (\cref{subsec:PCM-opt}).
We will use the notation in \cref{def:parallel-X-rotation} to represent a set of parallel $X$ gates, where a binary string specifies the qubits to which the $X$ gates are applied.
\begin{definition}[Parallel $X$-rotation]
$X_s$, where $s$ is a binary string of length $|s| = l$, denotes a $l$-qubit operation constructed as:
$ 1_{s_0}(X_0) \otimes \dots \otimes 1_{s_{l-1}}(X_{l-1})$,
where $1_{s_i}(X_i)$ denotes applying $X$-rotation to $i$-th qubit if the $i$-th bit of $s$ is $1$.
\label{def:parallel-X-rotation}
\end{definition}

\begin{example}
    $X_{101}$ denotes a circuit operator where on qubits $0$ and $2$ an $X$-rotation is applied, because the binary string $101$ is $1$ on bits $0$ and $2$:
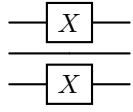
\begin{equation*}
    \begin{quantikz}[row sep = 0.15cm]
        & \gate{X} & \\
        &          & \\
        & \gate{X} & 
\end{quantikz}
\end{equation*}
    
\end{example}


We will use the notation in \cref{def:state-transform-circuit} to denote circuits that take one specific state to another specific state. See \cref{ex:state-transform-1}.
\begin{definition}[State transformation circuit]
For two $n$-qubit states $\ket{\psi}$, and $\ket{\phi}$, $C^{\ket{\phi}}_{\ket{\psi}} = \{C \mid  C\ket{\psi} =\ket{\phi}\} $. 
\label{def:state-transform-circuit}
\end{definition}
\begin{example}
    $X \in C^{\ket{0}}_{\ket{1}}$, because $X\ket{1} = \ket{0}$; $H\otimes H \in C^{\ket{00}}_{\ket{+-}}$, since $(H\otimes XH)\ket{+-} = (H\ket{+})\otimes(XH\ket{-})=\ket{0} \otimes (X\ket{1}) = \ket{0}\otimes\ket{0} =\ket{00}$.
    \label{ex:state-transform-1}
\end{example}


With \cref{lemma:state-transform-circ}, we demonstrate that for any pair of $n$-qubit states, it is straightforward to construct a circuit that transforms one state into the other, provided access to a state preparation procedure.

\begin{lemma}
Given two $n$-qubit states $\ket{\psi}$, $\ket{\phi}$, define $T_{\ket{\psi} \rightarrow \ket{\phi}} := SP_{\ket{\phi}}(SP_{\ket{\psi}})^{-1}$, where the definition of $SP$ is \cref{def:state-prep}, and for any circuit $C$, $C^{-1}$ denotes the inverse circuit of $C$, then it holds that $T_{\ket{\psi} \rightarrow \ket{\phi}} \in C^{\ket{\phi}}_{\ket{\psi}}$, and the circuit depth and gate count of $T_{\ket{\psi} \rightarrow \ket{\phi}}$ is bounded by $\mathcal{O}(2^n)$.
    \label{lemma:state-transform-circ}
\end{lemma}

\cref{theorem:measurement-on-entanglement} and \cref{theorem:reset-on-entanglement} form the foundation of our optimization strategy by establishing a systematic method to replace mid-circuit measurements or resets with probabilistic sub-circuits or static sub-circuits. 
In specific, \cref{theorem:measurement-on-entanglement} provides a way to replace mid-circuit measurements with probabilistic sub-circuits that emulate the probabilistic behavior of measurements. This replacement eliminates the need for classical feedback and synchronization during circuit execution, allowing the quantum computation to proceed uninterrupted.

\begin{theorem}
    Given an $n$-qubit state $\ket{\psi} \coloneqq \alpha_{0\dots 0}\ket{0\dots 0} + \dots+ \alpha_{1\dots 1}\ket{1\dots 1}$, 
    an $n$-qubit static circuit $T_{\ket{\psi}\rightarrow\ket{0}^{\otimes n}}$ which is defined in \cref{lemma:state-transform-circ}, then \cref{equation:theorem-measurement-on-entanglement} holds, where $X_s$ in which $s$ is a binary string of length $|s| = n$ is an $n$-qubit circuit defined in \cref{def:parallel-X-rotation}, and $\triangleq$ is the runtime equivalence defined in \cite{chen2024reducingmidcircuitmeasurementsprobabilistic}.
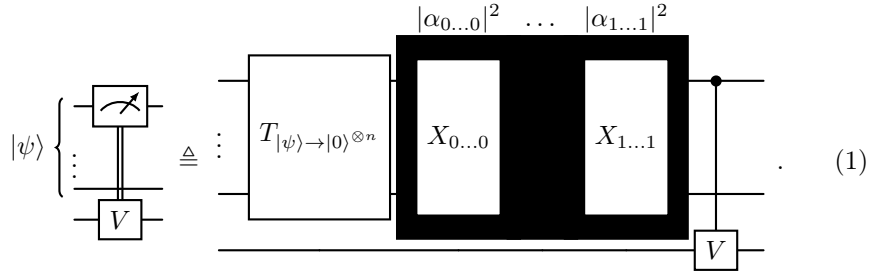
        \begin{align}
             \begin{quantikz}[row sep=0.15cm, column sep= 0.2cm]
            \lstick[3]{$\ket{\psi}$}  & \meter{} \wire[d][3]{c}& \\
            \vdots \\
                        &                        & \\
                        & \gate{V}                & 
            \end{quantikz}
            \triangleq 
             \begin{quantikz}[row sep=0.15cm, column sep= 0.35cm]
            &\gate[3]{T_{\ket{\psi}\rightarrow\ket{0}^{\otimes n}}}&\gate[3]{X_{0\dots0}}\gategroup[3,steps=1,style={fill=black, inner
     sep=3.7pt}, background]{$|\alpha_{0\dots0}|^2$}& \gategroup[3,steps=1,style={fill=black, inner
     sep=3.7pt}, background]{$\dots$} &\gate[3]{X_{1\dots1}}\gategroup[3,steps=1,style={fill=black, inner
     sep=3.7pt}, background]{$|\alpha_{1\dots1}|^2$}&\ctrl{3}  &\\
     \vdots \\
            &             && \ldots &&                       & \\
            &             &&&& \gate{V}                & 
            \end{quantikz}
            \,.
            \label{equation:theorem-measurement-on-entanglement}
        \end{align}\
    \label{theorem:measurement-on-entanglement}
\end{theorem}

In contrast to mid-circuit measurements, mid-circuit resets—where a qubit is deterministically initialized to a $\ket{0}$—do not involve probabilistic behaviors. \cref{theorem:reset-on-entanglement} defines the conditions under which the reset of an entangled qubit can be replaced by a sequence of circuits, ensuring the same final state without introducing additional quantum runtime overhead or probabilistic behavior.
\begin{theorem}
    Given an $n$-qubit state $\ket{\psi} \coloneqq \alpha_{0\dots 0}\ket{0\dots 0} + \dots+ \alpha_{1\dots 1}\ket{1\dots 1}$, an $n$-qubit circuits $T_{\ket{\psi}\rightarrow\ket{\phi}}$  that is defined in \cref{lemma:state-transform-circ}, where $\ket{\phi}$ is the state after resetting the qubit $q_0$ to $0$, then \cref{equation:theorem-reset-on-entanglement} holds.
        \begin{align}
             \begin{quantikz}[row sep=0.15cm, column sep= 0.2cm]
            \lstick[3]{$\ket{\psi}$}  & \push{\ket{0}} & \rstick{$q_0$} \\
            \vdots \\
                        &                        &
            \end{quantikz}
            \triangleq 
             \begin{quantikz}[row sep=0.15cm, column sep= 0.35cm]
            &\gate[3]{T_{\ket{\psi}\rightarrow\ket{\phi}}} &\\
     \vdots \\
            &                       & 
            \end{quantikz}
            \,.
            \label{equation:theorem-reset-on-entanglement}
        \end{align}\
    \label{theorem:reset-on-entanglement}
\end{theorem}


\paragraph{Optimization framework}
Our optimization framework is presented in \cref{alg:optimization}. This framework uses QCP to collect compile-time information on states, and replaces mid-circuit measurements and resets by \cref{theorem:measurement-on-entanglement} and \cref{theorem:reset-on-entanglement} taking advantage of the information gathered. In \cref{alg:optimization}, we introduce the parameter $n_{pcm}$, which serves as a critical control for the optimization process. 
The choice of $n_{pcm}$ governs the trade-off between circuit synthesis complexity and optimization power. A larger $n_{pcm}$ enables the framework to handle larger quantum states during optimization, thereby replacing more mid-circuit measurements and resets. This can result in a more optimized circuit. However, the increased $n_{pcm}$ comes at a cost:
\begin{itemize}
    \item \textbf{Time overhead:} Synthesis for quantum states of larger size (\cref{def:size-state}) requires more computational time.
    \item \textbf{Circuit complexity:} The generated circuits for states of larger size have higher depths and gate counts.
\end{itemize}
By tuning $n_{pcm}$, the user can balance these trade-offs, tailoring the optimization framework to the specific requirements of the application, such as minimizing quantum runtime or conserving resources.

\begin{algorithm}[htb]
\caption{Overall framework of optimization}
\label{alg:overall_framework}
    \KwData{$C \in dynamic \ circuits$, $n_{pcm} \ge 1$} 
    
    \KwResult{$C_o$} 
    $(C_o, \text{S}_{\text{const\_info}})  \gets (C, \text{QCP}.\textbf{run}(C))$\;
    
    \For {mid-circuit measurement $M$ in $C_o$}{
        
        \If{$\top \notin M.\text{input-state } \land n_{pcm} \ge |M.\text{input-state}|$ }{ \Comment{The state to be measured is totally known at compile-time,\\}
                \Comment{and the size of state (\cref{def:size-state}) is not larger than $n_{pcm}$. \\}
                replace $M$ by \cref{theorem:measurement-on-entanglement}\;

        }
    }

    \For {mid-circuit reset $R$ in $C_o$ }{
        
        \If{$\top \notin R.\text{input-state } \land n_{pcm} \ge |R.\text{input-state}|$}{ 
                replace $R$ by \cref{theorem:reset-on-entanglement}\;
        }
    }
    \Return $C_o$ \;
\label{alg:optimization}
\end{algorithm}

\subsection{Improvement on QCP}\label{subsec:ext-QCP}

\paragraph{Separation of unnecessary entanglements} 
When an entanglement forms, the entanglement groups (the entanglement group is a dedicated data structure in the union-table of QCP) of the affected qubits are combined. In the original implementation of QCP, however, even if, at some point, some qubits get disentangled from an entanglement after executing some instruction, their state remains grouped inside the same entanglement group, significantly hindering the efficiency of state representation.
Take the following $3$-qubit circuit for example:
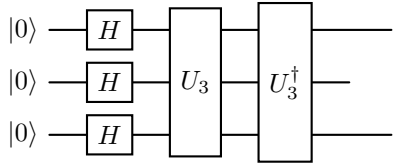
\begin{equation*}
\begin{quantikz}[row sep=0.08cm]
    \lstick{\ket{0}}&\gate{H}&\gate[3]{U_3}&\gate[3]{U_3^{\dagger}}&\\
    \lstick{\ket{0}}&\gate{H}&&&\\
    \lstick{\ket{0}}&\gate{H}&&&
\end{quantikz}
\end{equation*}
In the previous version of QCP, after executing the circuit, the state information in the union table is stored in the entanglement group $\{|000\rangle \to \frac{1}{2\sqrt{2}}, |001\rangle \to \frac{1}{2\sqrt{2}},|010\rangle \to \frac{1}{2\sqrt{2}},|011\rangle \to \frac{1}{2\sqrt{2}},|100\rangle \to \frac{1}{2\sqrt{2}},|101\rangle \to \frac{1}{2\sqrt{2}},|110\rangle \to \frac{1}{2\sqrt{2}},|111\rangle\}$, where the $3$-qubit state remains in the same storage group, even if any pair qubits is no longer entangled. 
In this work, a new feature is introduced to separate disentangled qubits from unnecessary storage groups.
With our current implementation, the state information after executing the circuit above is stored in $3$ separate entanglement groups of the same configuration $\{\ket{0} \to \frac{1}{\sqrt{2}}\}$.
With this feature, the growth of entanglement groups slows down, reducing the size of entries maintained by the union-table. Consequently, it allows for our implementation to track more state information than the original QCP at compile-time.

\paragraph{Propagation through resets} For a state of the entire system $\ket{\psi}$, we denote its abstract state in QCP by $s$. In the previous version of QCP, $[\![reset_i]\!]^{\sharp}s = \top$, where $[\![reset_i]\!]^{\sharp}$ is the abstract effect of resetting $i$-th qubit to $\ket{0}$. It means whenever a qubit in a state known by QCP is reset, all information on this state will be lost afterwards. This significantly limits QCP's capability to track information. We improve QCP by allowing information propagation on the state $s$ after a reset operation. In the current version, suppose that $s = \sum_j \lambda_j\ket{\psi^j}$, then $[\![reset_i]\!]^{\sharp}s = \frac{1}{\lambda_{norm}} (\sum_{\psi^j_i = 0} \lambda_j \ket{\psi^j})$, where $\psi_i^j$ indicates the $i$-th qubit of the basis state $\ket{\psi^j}$ and $\lambda_{norm}$ is used to normalize the state after the reset.

\section{Evaluation}\label{sec:evaluation}
\subsection{Experiments}
Qiskit provides several optimization passes for compiling quantum circuits, two of which are particularly relevant for comparison with our approach:
\begin{itemize} \item $\texttt{RemoveResetInZeroState}$ eliminates resets when the qubit is in $\ket{0}$.
\item $\texttt{ResetAfterMeasureSimplification}$ replaces resets occurring after measurements with a conditional X gate.
\end{itemize}

Specifically, we first evaluated our approach with $n_{pcm} = 1$, which corresponds to the PCM-based optimization pass proposed in \cite{chen2024reducingmidcircuitmeasurementsprobabilistic}, to provide a baseline comparison. We then progressively increase $n_{pcm}$ to investigate the impact of higher optimization levels on mid-circuit measurement and reset removal, execution time, and the additional operations introduced as replacements (\cref{theorem:measurement-on-entanglement}, \cref{theorem:reset-on-entanglement}). Additionally, we analyzed the trade-offs between optimization effectiveness and computational overhead as $n_{pcm}$ increased, as discussed in \cref{subsec:improved-PCM-pass}.
The experiments were conducted on a machine equipped with an Apple M2 Pro chip and 16GB of RAM, and results are presented in \cref{subsec:results}.
\subsection{Dataset}\label{subsec:dataset}
We generated our dataset using a modified version of the $\texttt{random\_circuit()}$ function provided by Qiskit \cite{ibmRandomcircuitv019}, to examine the behavior of our optimization pass across a range of scenarios with circuits of varying sizes (i.e., qubit number and depths). Beyond what is offered by $\texttt{random\_circuit()}$, we include a parameter that controls the density of mid-circuit measurement operations generated and a feature to generate sub-circuits consisting of operations that are controlled by classical bits.
For our experiments, we generated quantum circuits of varying sizes, with the integer parameter $scale$ determining both the number of qubits and the circuit depth. Specifically, the number of qubits was set to $scale \times 10$, and the circuit depth to $scale \times 200$. For example, when $scale = 3$, the generated circuit consists of a circuit with $30$ qubits and a depth of $600$.

\subsection{Results}
\label{subsec:results}
\begin{figure}[h]
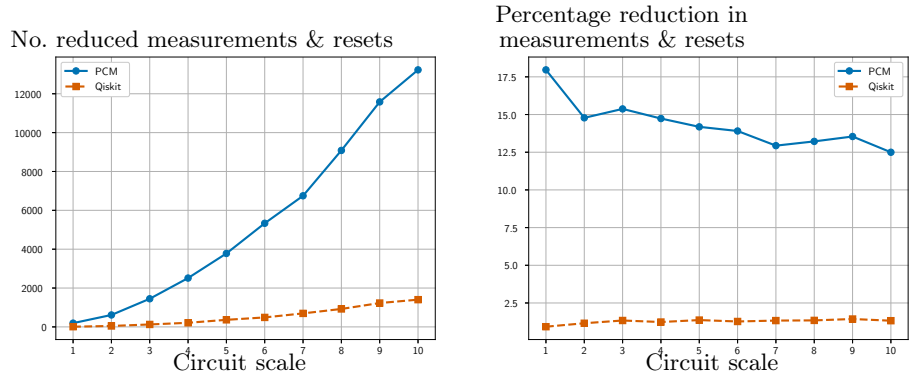

    \centering
    \begin{subfigure}[b]{0.485\textwidth} 
        \centering
        \begin{tikzpicture}[>=latex,node distance=0.3em]
         \node(a)
         {
        \resizebox{1.1\textwidth}{!}{\input{images/scale_num.pgf}}
         };
         \node[] at (0,-2.2)
        {\text{Circuit scale}};         \node[] at (-0.5,2.1)
        {\text{No. reduced measurements \& resets}};
        \end{tikzpicture}
        \label{fig:scale_num}
    \end{subfigure}
    \hfill
    \begin{subfigure}[b]{0.485\textwidth} 
        \centering
        \begin{tikzpicture}[>=latex,node distance=0.1em]
         \node(a)
         {
        \resizebox{1.1\textwidth}{!}{\input{images/scale_perc.pgf}}
         };
         \node[] at (0,-2.2)
        {\text{Circuit scale}};
        \node[] at (-1.2,2.4)
        {\text{Percentage reduction in}};
        \node[] at (-1.2,2.1)
        {\text{measurements \& resets}};
        \end{tikzpicture}
        \label{fig:scale_perc}
    \end{subfigure}

    \caption{Performance comparison between PCM-based optimization framework with $n_{pcm} = 1$ and Qiskit in terms of mid-circuit measurement and reset removals as the circuit scale (see \cref{subsec:dataset}) increases. 
    }
    \label{fig:scale}
\end{figure}

The data presented in the plots for each circuit scale (defined in \cref{subsec:dataset}) are obtained by  evaluating $10$ quantum circuits of the corresponding scale and summing up the results.
\cref{fig:scale} illustrates the comparison between our optimization framework with $n_{pcm} = 1$ and Qiskit in terms of their performance in reducing mid-circuit measurements and resets. The results clearly demonstrate that our optimization framework significantly outperforms Qiskit, even when $n_{pcm} = 1$, achieving removal rates that surpass $17\%$. In contrast, Qiskit's removal rates remain notably lower, consistently under $2\%$. This difference highlights superior efficiency of our PCM-based optimization framework in removing mid-circuit measurements and resets.

\begin{figure}[htb]
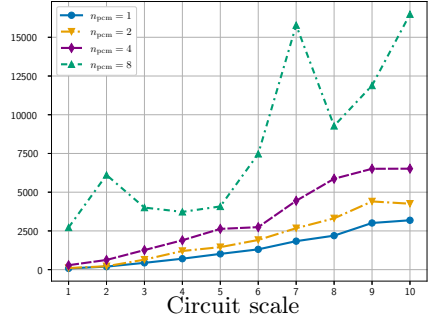
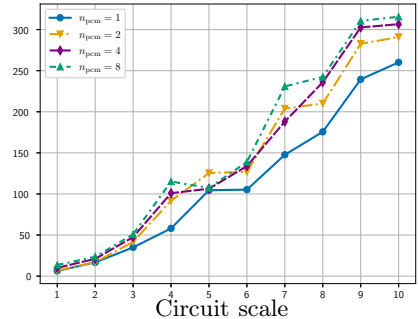
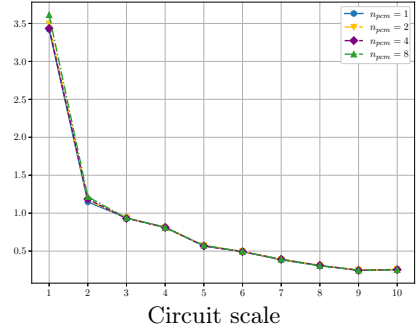

    \centering
    \begin{subfigure}[b]{0.48\textwidth}
        \centering
        \begin{tikzpicture}[>=latex,node distance=0.3em]
         \node(a)
         {
        \resizebox{1.1\textwidth}{!}{\input{images/comp_ext_num.pgf}}
         };
         \node[] at (0,-2.2)
        {\text{Circuit scale}};         \node[] at (-0.8,2.1)
        {\text{No. measurements \& resets}};
        \end{tikzpicture}
        \caption{}
        \label{fig:ext_num}
    \end{subfigure}
    \hfill
    \begin{subfigure}[b]{0.48\textwidth}
        \centering
         \begin{tikzpicture}[>=latex,node distance=0.3em]
         \node(a)
         {
        \resizebox{1.1\textwidth}{!}{\input{images/comp_ext_extr.pgf}}
         };
         \node[] at (0,-2.2)
        {\text{Circuit scale}};         \node[] at (-1.4,2.1)
        {\text{No. introduced gates}};
        \end{tikzpicture}
        \caption{}
        \label{fig:ext_extr}
    \end{subfigure}

    \vspace{1em}

    \begin{subfigure}[b]{0.48\textwidth}
        \centering
         \begin{tikzpicture}[>=latex,node distance=0.3em]
         \node(a)
         {
        \resizebox{1.1\textwidth}{!}{\input{images/comp_ext_time.pgf}}
         };
         \node[] at (0,-2.2)
        {\text{Circuit scale}};         \node[] at (-0.8,2.1)
        {\text{Time consumed (in seconds)}};
        \end{tikzpicture}
        \caption{}
        \label{fig:ext_time}
    \end{subfigure}
    \hfill
    \begin{subfigure}[b]{0.48\textwidth}
        \centering
         \begin{tikzpicture}[>=latex,node distance=0.3em]
         \node(a) at (0.9, 0.2)
         {
        \resizebox{0.955\textwidth}{!}{\input{images/stand_dev.pgf}}
         };
         \node[] at (0.9,-2)
        {\text{Circuit scale}};         \node[] at (0.1,2.35)
        {\text{SD of removal percentages}};
        \end{tikzpicture}
        \caption{}
        \label{fig:stand_dev}
    \end{subfigure}

    \caption{Comparison of PCM performance under different $n_{pcm}$ settings. 
    }
    \label{fig:ext}
\end{figure}

Figure \cref{fig:ext} shows the performance of our optimization framework as $n_{pcm}$ increases. 
Specifically, \cref{fig:ext_num} shows the difference between the number of measurements and resets eliminated when using $n_{pcm} = 1$ (the lowest optimization level) and when using higher values of $n_{pcm}$. 
The results indicate that increasing optimization level $n_{pcm}$ allows our framework to be more aggressive and reduce more mid-circuit measurements and resets. However, as $n_{pcm}$ continues to increase, the additional reduction in mid-circuit measurements and resets becomes less significant compared to the gains achieved at lower values of $n_{pcm}$. This diminishing effect stems from the upper limit of QCP on the size of entanglement groups it can track. Specifically, the constraint imposed by the QCP parameter $n_{max}$ restricts the impact of PCM, regardless of how large $n_{pcm}$ becomes. 

By \cref{theorem:measurement-on-entanglement} and \cref{theorem:reset-on-entanglement},   additional static operations (see \cref{subsec:static-dynamic-circuit}) are introduced to replace mid-circuit measurements and resets. As shown in \cref{fig:ext_extr}, the number of introduced static operations increases as $n_{pcm}$ grows. This highlights the trade-off offered by our framework: by adjusting $n_{pcm}$, users can balance between minimizing dynamic operations (see \cref{subsec:static-dynamic-circuit}) that are high in runtime overhead and controlling the additionally introduced static overhead. 

\cref{fig:ext_time} reports the execution time consumed by state preparations in the optimization framework under different settings of $n_{pcm}$. It demonstrates that the computational cost rises as $n_{pcm}$ increases. This further confirms the flexibility of our framework: it allows users to tune $n_{pcm}$ to achieve the desired balance between optimization strength and computational efficiency.

Lastly, \cref{fig:stand_dev} reports the \emph{standard deviation} (SD) of the percentages of removed mid-circuit measurement and reset operations for different values of $n_{pcm}$. To obtain a reliable estimate of the standard deviation, we evaluated $50$ random circuits for each circuit scale. As shown in the figure, the standard deviation is relatively high at the smallest scale, indicating greater variability in optimization outcomes for small circuits. However, the SD quickly decreases as the circuit scale increases, stabilizing below $0.5$ from scale $6$ onwards across all values of $n_{pcm}$. This trend suggests that the effectiveness of our optimization framework becomes increasingly consistent as the circuit size grows. 
 
\subsection{Demonstrative example: optimizing a quantum algorithm}
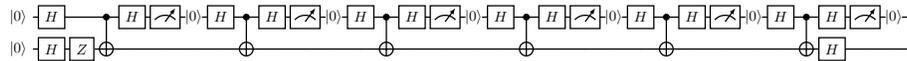
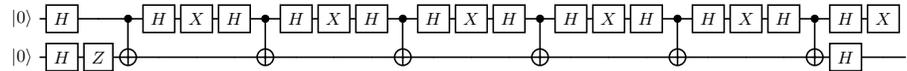
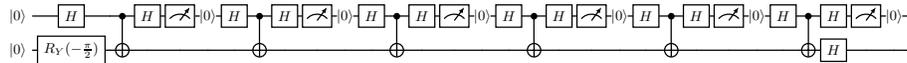
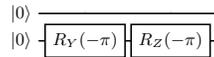
\begin{figure}[htb]
    \centering
    \begin{subfigure}[b]{1\textwidth}
        \centering
        \resizebox{1\textwidth}{!}{  
        \begin{quantikz}[row sep=0.2cm, column sep=0.11cm]
            \lstick{$\ket{0}$} & \gate{H} & & \ctrl{1} & \gate{H} & \meter{} & \push{\ket{0}} & \gate{H} & \ctrl{1} & \gate{H} & \meter{} & \push{\ket{0}} & \gate{H} & \ctrl{1} & \gate{H} & \meter{} & \push{\ket{0}} & \gate{H} & \ctrl{1} & \gate{H} & \meter{} & \push{\ket{0}} & \gate{H} & \ctrl{1} & \gate{H} & \meter{} & \push{\ket{0}} & \gate{H} & \ctrl{1} & \gate{H} & \meter{} & \push{\ket{0}} & \\
            \lstick{$\ket{0}$} & \gate{H} & \gate{Z} & \targ{} & & & & & \targ{} & & & & & \targ{} & & & & & \targ{} & & & & & \targ{} & & & & & \targ{} & \gate{H} & & &
        \end{quantikz}
        }
        \caption{The circuit of Bernstein-Vazirani algorithm with qubit reuse employed.}
        \label{fig:dem_exmp_before}
    \end{subfigure}
    \hfill
    
    \begin{subfigure}[b]{1\textwidth}
        \centering
    \resizebox{1\textwidth}{!}{  
        \begin{quantikz}[row sep=0.2cm, column sep=0.11cm]
            \lstick{$\ket{0}$} & \gate{H} & & \ctrl{1} & \gate{H} & \gate{X} & \gate{H} & \ctrl{1} & \gate{H} & \gate{X} & \gate{H} & \ctrl{1} & \gate{H} & \gate{X} & \gate{H} & \ctrl{1} & \gate{H} & \gate{X} & \gate{H} & \ctrl{1} & \gate{H} & \gate{X} & \gate{H} & \ctrl{1} & \gate{H} & \gate{X} \\
            \lstick{$\ket{0}$} & \gate{H} & \gate{Z} & \targ{} & & & & \targ{} & & & & \targ{} & & & & \targ{} & & & & \targ{} & & & & \targ{} & \gate{H} & & 
        \end{quantikz}
    }
        \caption{Optimized circuit by our PCM-based optimization framework.}
        \label{fig:dem_exmp_after}
    \end{subfigure}

    \hfill
    \begin{subfigure}[b]{1\textwidth}
        \centering
    \resizebox{1\textwidth}{!}{  
        \begin{quantikz}[row sep=0.2cm, column sep=0.11cm]
            \lstick{$\ket{0}$} & \gate{H} & & \ctrl{1} & \gate{H} & \meter{} & \push{\ket{0}} & \gate{H} & \ctrl{1} & \gate{H} & \meter{} & \push{\ket{0}} & \gate{H} & \ctrl{1} & \gate{H} & \meter{} & \push{\ket{0}} & \gate{H} & \ctrl{1} & \gate{H} & \meter{} & \push{\ket{0}} & \gate{H} & \ctrl{1} & \gate{H} & \meter{} & \push{\ket{0}} & \gate{H} & \ctrl{1} & \gate{H} & \meter{} & \push{\ket{0}} & \\
            \lstick{$\ket{0}$} & \gate{R_Y(-\frac{\pi}{2})} &  & \targ{} & & & & & \targ{} & & & & & \targ{} & & & & & \targ{} & & & & & \targ{} & & & & & \targ{} & \gate{H} & & &
        \end{quantikz}
        }
        \caption{Circuit obtained by applying Qiskit's transpiler to \cref{fig:dem_exmp_before}.}
        \label{fig:dem_exmp_after_just_qiskit}
    \end{subfigure}
    \hfill
    
    \begin{subfigure}[b]{1\textwidth}
        \centering
    \resizebox{0.25\textwidth}{!}{  
        \begin{quantikz}[row sep=0.2cm, column sep=0.11cm]
            \lstick{$\ket{0}$} & & & \\
            \lstick{$\ket{0}$} & \gate{R_Y(-\pi)} & \gate{R_Z(-\pi)} & 
        \end{quantikz}
    }
        \caption{Circuit obtained by applying Qiskit's transpiler to \cref{fig:dem_exmp_after}.}
        \label{fig:dem_exmp_after_transp}
    \end{subfigure}
    
    \caption{Dimonstrative example of applying the PCM-based optimization framework the circuit of Bernstein-Vazirani algorithm employing qubit reuse.}
    \label{fig:dem_exmp}
\end{figure}

We demonstrate the effectiveness of our method on a canonical quantum algorithm: the Bernstein-Vazirani algorithm \cite{Bernstein-Vazirani}. Specifically, we consider a version of this quantum circuit implemented using the \textit{qubit reuse} technique \cite{Bernstein-Vazirani-Reuse}. Qubit reuse involves resetting and reinitializing a qubit after measurement, enabling its reuse within the same quantum circuit and thereby saving hardware resources.
After applying our optimization framework to \cref{fig:dem_exmp_before}, we get an optimized circuit \cref{fig:dem_exmp_after} with much lower runtime overhead, where all resets are reduced.

To further demonstrate the influence of our optimization, we apply Qiskit’s transpiler to both \cref{fig:dem_exmp_before} and \cref{fig:dem_exmp_after}, and we obtain circuits \cref{fig:dem_exmp_after_just_qiskit} and \cref{fig:dem_exmp_after_transp}, respectively. We observe that while resets in \cref{fig:dem_exmp_before} limit Qiskit’s ability to optimize the circuit structure, our framework eliminates these resets, unlocking additional optimization opportunities for Qiskit’s compilation passes. This sequential approach—first applying our optimization to remove resets and then leveraging Qiskit’s optimization—produces a highly optimized circuit, demonstrating how our framework can complement existing compilation toolchains.


\section{Conclusion and future works}\label{sec:conclusion}
In this work, we developed a PCM-based optimization framework to reduce mid-circuit measurements and resets in quantum circuits. By incorporating circuit synthesis, our approach extends optimization to measurements and resets on entangled states. The framework introduces a tunable parameter $n_{pcm}$, allowing a trade-off between optimization level and computational cost. Our experimental results confirm that our method effectively reduces mid-circuit measurements and resets, significantly mitigating runtime overhead. Future work will focus on deeper integration with compilation toolchains and extending optimizations for broader classes of dynamic circuits.




\begin{credits}
\subsubsection{\ackname} We are grateful to Prof. Dr. Helmut Seidl for many fruitful discussions and his support at all times. This work has been supported by the Bavarian state government through the project Munich Quantum Valley with funds from the Hightech Agenda Bayern Plus.

\subsubsection{\discintname}
The authors have no competing interests to declare that are relevant to the content of this article. 
\end{credits}
%
%
%
\bibliographystyle{splncs04}

%




\end{document}